# Magneto-Gravitational Regulated Streamer Accretion onto a Class 0 Protostellar System

Bo Huang[1,2]*, Josep M. Girart[1,3], Ian W. Stephens[4], Tom Megeath[5], Valentin J. M. Le Gouellec[1,3], Nadia M. Murillo[6], Paulo Cortés[7,8], Manuel Fernández-López[1,9,10], Zhi-Yun Li[11], Leslie W. Looney[12], J. A. López-Vázquez[13], Jaime E. Pineda[14], Álvaro Sánchez-Monge[1,3], Patricio Sanhueza[15], Sarah Sadavoy[16], Qizhou Zhang[17], Charles L. H. Hull[18,19], Nicole Karnath[17,20], Enwei Liang[21], Philip C. Myers[17]

[1]Institut de Ciències de l'Espai (ICE-CSIC); Cerdanyola del Vallés, E-08193, Catalonia, Spain.
[2]Korea Astronomy and Space Science Institute; Daejeon, 34055, Republic of Korea
[3]Institut d'Estudis Espacials de Catalunya (IEEC); Castelldefels, E-08860, Catalonia, Spain.
[4]Department of Earth, Environment, and Physics, Worcester State University; Worcester, MA 01602, USA.
[5]Ritter Astrophysical Research Center, Department of Physics and Astronomy, University of Toledo; Toledo, OH 43606, USA.
[6]Instituto de Astronomía, Universidad Nacional Autónoma de México; Ensenada, CP 22830, México.
[7]National Radio Astronomy Observatory; Charlottesville, VA 22093, USA.
[8]Joint ALMA Observatory; Santiago, Chile.
[9]Instituto Argentino de Radioastronomía; Buenos Aires, 1888, Argentina.
[10]Facultad de Ciencias Astronómicas y Geofísicas, Universidad Nacional de La Plata; Paseo del Bosque S/N, B1900FWA La Plata, Argentina.
[11]Astronomy Department, University of Virginia; Charlottesville, VA 22904, USA.
[12]Department of Astronomy, University of Illinois; Urbana, IL 61801, USA.
[13]Academia Sinica Institute of Astronomy and Astrophysics; Taipei, 10617, Taiwan.
[14]Center for Astrochemical Studies, Max Planck Institute for Extraterrestrial Physics; Garching, D-85748, Germany.
[15]Department of Astronomy, The University of Tokyo; Tokyo, 113-8654, Japan.
[16]Department of Physics, Queen's University; Kingston, ON K7L 3N6, Canada.
[17]Center for Astrophysics | Harvard & Smithsonian; Cambridge, MA 02138, USA.
[18]AES Indiana; Indianapolis, IN 46204, USA.
[19]National Astronomical Observatory of Japan; Santiago, Chile.
[20]Space Science Institute; Suite B Boulder, CO 80301, USA.
[21]School of Physical Science and Technology, Guangxi University; Nanning, 530004, China.

*Corresponding to: huang@ice.csic.es

## Abstract

How do magnetic fields shape the way young stars gather gas from their birth clouds? Using high-resolution Atacama Large Millimeter/submillimeter Array observations of a young triple protostellar system HOPS-182, we identify an elongated stream of gas, or accretion streamer, that extends over several thousand astronomical units (1 astronomical unit is the Earth-Sun distance) and carries a substantial flow of material toward the system. The gas speeds along this filament increase toward the star in a way consistent with gravitational free-fall, while the streamer's shape closely follows the magnetic field threading the region. By comparing the strengths of gravity and magnetic tension and measuring how the gas rotates compared with the local magnetic field, we show that the field is strong enough to help confine and guide the infalling gas and efficiently remove angular momentum. These results suggest that a substantial fraction of the material falling onto young protostellar systems can be funneled through elongated, magnetically structured accretion streamers.

**Teaser**

Magnetic fields help funnel gas along the accretion streamer that feed the growth of a young protostellar system.

# MAIN TEXT

# Introduction

Star formation involves the transfer of mass from large-scale cloud material onto compact protostars and disks. Observations reveal that this transfer often occurs through narrow, asymmetric filamentary structures—commonly referred to as accretion streamers—which channel material from envelopes, typically spanning scales of ~$10^3$–$10^4$ astronomical units (AU, 1 AU = 149,597,870.7 km) down to disk scales of ~$10^2$ AU (*1*–*5*). These streamers are seen across various protostellar environments and are thought to play a critical role in sustaining disk growth and stellar mass accumulation. Idealized models of magnetized collapse predict that magnetic fields tend to align with infalling flows and exhibit pinched morphologies (*6*–*8*). However, it remains unclear whether accretion flows are dynamically guided by magnetic fields, primarily follow gravitational potential gradients, or reflect a competition between these forces (*9*). Resolving this question requires spatially resolved measurements of both the gas kinematics and the magnetic field structure within streamer-bearing systems.

Recently, we presented the largest polarization and magnetic field study from the *B*-field Orion Protostellar Survey (BOPS), encompassing 61 protostars within the Orion Molecular Cloud (OMC) complex (*10*, *11*). Among this sample, six sources exhibit magnetic fields aligned with elongated dust structures, with HOPS-182 standing out as a particularly striking example (*11*). As shown in Fig. 1, which presents the 0.87 mm (345 GHz) dust emission from different tracers observed with Atacama Large Millimeter/submillimeter Array (ALMA) with an angular resolution of ~0.7" (~280 AU), HOPS-182 drives a large-scale (>10,000 AU) and high-velocity (>50 km $s^{-1}$) bipolar outflow—a common feature in young protostars where mass is transported along the poles due to angular momentum conservation (*12*). The velocity field, traced by the optically thin molecular line of $C^{17}O$ (3–2) (*11*) at 0.87 km $s^{-1}$ spectral resolution, is spatially extended and exhibits a clear velocity gradient along the southern elongated dust structure. The magnetic field morphology, inferred from polarized dust emission due to magnetically aligned dust grains (*13*, *14*), is broadly aligned with this southern elongated dust structure (see Fig. 1B). HOPS-182 is a triple Class 0 (young protostellar phase) protostellar system (*15*) at a distance of ~390 parsec (pc, 1 pc = 206,264.8 AU) from Earth (*16*), with the two central components separated by ~390 AU and resided within regions that are considered to be “potential disks”, while a third, fainter source lies ~2000 AU to the southeast and contributes only ~2% of the 0.87 mm continuum peak flux (*11*), suggesting minimal dynamical impact on the accretion environment. Besides, HOPS-182 has a relatively high luminosity of ~71 $L_\odot$ (where $L_\odot$ represents the solar luminosity) and a temperature of ~51.9 K, indicative of a deeply embedded and actively accreting system (*15*, *17*). Radiative transfer modeling of its spectral energy distribution (SED) yields an envelope mass of ~0.33 $M_\odot$ (where $M_\odot$ represents the solar mass) and an envelope mass infall rate of ~$2.5 \times 10^{-6}$ $M_\odot$ year$^{-1}$ (*18*). Guided by the source luminosity, the envelope mass, and the accretion rate, we adopt $0.8 \leq M_* \leq 2.0$ $M_\odot$ as the allowed range of total stellar masses embedded in the central beam (which includes both stars and the inner envelope around the dust peak intensity) (*10*).

# Results and Discussion

### Kinematics and the accretion rate

We used ancillary 1.36 mm (220 GHz) ALMA observations that include high-spectral-resolution (~0.04 km s$^{-1}$) $C^{18}O$ (2–1) line emission with an angular resolution of ~1.2" (~460 AU), enabling a more detailed assessment of the kinematics than provided by the BOPS data (*10*). The 1.36 mm images (see panels A and B in Fig. 2) reveal a structure similar to that seen at 0.87 mm: the $C^{17}O$ (3–2) and $C^{18}O$ (2–1) line emissions trace extended structures with comparable spatial distributions and velocity ranges, the dust continuum emission shows similar elongated features, and the CO (2–1) and SiO (5–4) transitions reveal outflow components with consistent morphologies. Using this high-spectral-resolution data, we determine the observed streamer spine by computing intensity-weighted means of the right ascension (RA), declination (Dec), and radial velocity in ten bins defined by a distance metric (red points in Fig. 2B and 2D). We then use the TIPSY (Trajectory of Infalling Particles in Streamers around Young stars) framework (*19*), which implements the analytical solution for initially rotating infall from Mendoza's gravitational infall model (*20*), to fit this streamer spine in position-position-velocity (PPV) space. Within the stellar mass range of $0.8 \leq M_* \leq 2.0$ $M_\odot$, the best-fit streamer trajectories (black curves Fig. 2B–2D) remain essentially unchanged (*10*). We extracted a position-velocity (PV) diagram from the $C^{18}O$ (2–1) data cube along the best-fit TIPSY and the observed streamer trajectories to assess the kinematics of the streamer (Fig. 3). We find that these PV diagrams are indistinguishable from each other, and gas located farther from the protostar exhibits lower velocities, while gas at small offsets moves more rapidly along the trajectory. This observed velocity gradient for both trajectories is consistent with gravitational acceleration of material flowing inward the central protostellar system.

According to the TIPSY best-fit solutions, the corresponding timescale ranges from $(5.1 \pm 1.4) \times 10^4$ years at $M_* = 0.8$ $M_\odot$ to $(1.7 \pm 0.1) \times 10^4$ years at $M_* = 2.0$ $M_\odot$. Based on the 0.87 mm dust continuum emission along the streamer region defined by TIPSY, we derived a fitted streamer mass of ~$0.07 \pm 0.03$ $M_\odot$. This corresponds to a mass infall rate ranging from $(1.2 \pm 0.6) \times 10^{-6}$ to $(3.9 \pm 1.6) \times 10^{-6}$ $M_\odot$ year$^{-1}$ (*10*). The fitted accretion streamer contains a mass of ~16%–20% of the total envelope mass (~0.33 $M_\odot$ from SED modeling or ~0.42 $M_\odot$ from dust continuum emission) (*12*, *18*), while the derived infall rate is comparable to, or at least a substantial fraction of the envelope infall rate of ~$2.5 \times 10^{-6}$ $M_\odot$ year$^{-1}$ (*18*). This indicates that the streamer provides an efficient and potential episodic channel for mass accretion that can contribute substantially to the protostellar growth during specific evolutionary phases. In addition to the dominant southern streamer, we find another shorter elongated dust structure east of the protostar in the 0.87 mm continuum map (Fig. 1). The magnetic field appears to partially trace this structure, but the absence of kinematic information prevents confirmation of its dynamical nature. If this or similar features are also accreting, it would imply that mass delivery during the Class 0 phase occurs through multiple, anisotropic, and potentially magnetically influenced streamers, providing an efficient and organized alternative to isotropic envelope collapse.

## Magnetic field, gravity, and accretion streamer

From the starting point of the TIPSY fit, we found that in the first ~1000 AU, the magnetic field follows the streamer kinematics trajectory better than the TIPSY one. Indeed, the mean orientation angle difference between the streamer and the magnetic field trajectory is ~2°, and between the streamer and the TIPSY is ~ 10°. Closer to the protostars, the observed streamer and the magnetic field begin deviate from one to another (mean angle difference ~32°), while the streamer lies somewhat closer to the TIPSY trajectory (mean difference ~26°). This shift in relative alignment, from streamer-magnetic-field agreement at larger scales to closer streamer-TIPSY agreement in the inner region, suggests that the magnetic field is more influential in setting the flow geometry in the outer streamer, whereas gravity becomes increasingly dominant as the infalling gas moves inward. To investigate how these geometries evolve, we compare the magnetic field orientation with the local gravitational field direction, computed as the gradient of the gravitational potential inferred from the dust continuum emission, including the contribution from the stellar mass (*21*). The gravitational field is essentially unchanged across the stellar mass range, and the magnetic field aligns well with the gravitational field at the outer end of the southern structure (*10*). Closer to the protostar, the field and gravitational directions become more misaligned, suggesting a change in the relative influence of magnetic tension, gravity, and local gas dynamics. In the innermost region, the magnetic field appears to realign with the gravitational direction, consistent with a scenario in which field lines are partially dragged inward by the collapsing gas (*10*). These spatial varying alignments highlight a dynamic interplay between magnetic support and gravitational infall during the earliest stages of star formation.

To quantify the relative dynamical roles of magnetic fields and gravity in shaping the accretion streamer, we compare the magnetic tension force density $f_{mag}$ and the gravitational force density $f_{grav}$, defined as $f_{mag} = B^2\kappa/4\pi$ and $f_{grav} = GM_*\rho_{str}/D_{str}^2$ (*22*, *23*), where $B$ is the magnetic field strength, $\kappa$ is the field-line curvature, $\rho_{str}$ is the gas density, and $D_{str}$ is the distance to the protostar. For the fitted streamer we infer $B$ = 1.4 mG, $\kappa = (3.1 \pm 0.2) \times 10^{-17}$ cm$^{-1}$, $\rho_{str} = 9.5 \times 10^{-17}$ g cm$^{-3}$, and $D_{str} = (2.2 \pm 0.2) \times 10^3$ AU across the allowed range of stellar mass, respectively (*10*), which yield $f_{mag} = (0.5–2.1) \times 10^{-23}$ dyn cm$^{-3}$ when including a factor-of-2 uncertainty for $B$ (*24–26*), and $f_{grav}$ in a range $(0.5–3.1) \times 10^{-23}$ dyn cm$^{-3}$ including 40% uncertainty for $\rho_{str}$. These values indicate that magnetic tension and gravity are comparable in magnitude along the streamer, suggesting a magneto-gravitational state in which the magnetic field is dynamically important in supporting and shaping the flow, while gravity mainly drives the overall infall. At the position where the magnetic field curvature reaches its maximum, $\kappa_{max} = (5.0 \pm 0.3) \times 10^{-17}$ cm$^{-1}$, at a distance $D_{\kappa max} = (1.3 \pm 0.1) \times 10^3$ AU, we find $f_{mag,max} = (0.8–3.3) \times 10^{-23}$ dyn cm$^{-3}$ and $f_{grav,max} = (2.0–9.4) \times 10^{-23}$ dyn cm$^{-3}$ (*10*). The larger values of $f_{grav,max}$ at this inner location are consistent with an increasing gravitational pull that bends the field lines as the gas accelerates inward (*23*).

The elongated dust structure of the streamer may also reflect the influence of rotation. The balance between rotation and magnetic fields governs the ability of collapsing gas to form rotationally supported disks, with magnetic braking partially removing angular momentum from infalling material and thereby regulating disk formation and growth. A key diagnostic of this interplay is the ratio of angular velocity to magnetic field strength, $\omega/B$, which determines whether the dynamics are dominated by rotation or by magnetic regulation. If the measured $\omega/B$ exceeds the critical value $(\omega/B)_{crit} = 1.69 \times 10^{-7} \times (c_s / 0.19$ km s$^{-1})^{-1}$ year$^{-1}$ μG$^{-1}$ (*27*), where $c_s$ is the sound speed, centrifugal forces dominate the dynamics, otherwise magnetic forces regulate the dynamics. For an envelope gas temperature of 43.7 K with a 30% uncertainty (*10*), corresponding to $c_s = 0.40 \pm 0.06$ km s$^{-1}$, we derive $(\omega/B)_{crit} = (8.2 \pm 0.9) \times 10^{-8}$ year$^{-1}$ μG$^{-1}$. For the allowed range of stellar mass, the observed value $\omega/B = (0.5–3.6) \times 10^{-8}$ year$^{-1}$ μG$^{-1}$ (*10*) is at least two times smaller than the critical threshold, indicating that magnetic braking partially removes angular momentum from the infalling gas, thereby suppressing disk growth (*28*).

Consistently, the disk radius in HOPS-182 is ~40 AU, smaller than the BOPS Class 0 average disk radius of ~80 AU (*15*). Consequently, kinematic models of accretion streamers should eventually include the impact of magnetic braking, i.e., the reduction of angular momentum within the streamer.

In summary, our high-resolution ALMA observations of the Class 0 protostellar system HOPS-182 reveal a magneto-gravitational state of accretion streamer extending over thousands of AU, with a notable mass infall rate. The observed velocity gradient along the streamer is consistent with gravitational acceleration, while the near equality between $f_{mag}$ and $f_{grav}$ implies that magnetic forces contribute non-negligibly to the gas dynamics. In this regime, gravity mainly sets the overall infall, whereas magnetic tension structures the trajectory and helps maintain the narrow, elongated morphology of the streamer. Quantitative comparison of the angular-velocity-to-magnetic-field-strength ratio with the critical threshold further indicates that magnetic braking is important and that the field plays a dynamically important role in partially regulating angular-momentum transport. These results support a scenario in which accretion onto protostars can proceed through magneto-gravitational regulated streamers that efficiently deliver material onto the protostar.

# Materials and Methods

## Observations

The ALMA Band 6 data (220 GHz; project ID: 2018.1.01073.S; PI: Tom Megeath) were observed on 11 March 2019, reaching a spatial resolution of approximately 1.2" (corresponding to ~460 AU at a distance of 390 pc). Observations were performed in dual mode (XX, YY) using the frequency division mode (FDM) across seven spectral windows. The $C^{18}O$ (2–1) transition, with a rest frequency of 219.56 GHz, was included in the spectral setup and observed with a spectral resolution of ~0.04 km $s^{-1}$.

The ALMA Band 7 (345 GHz) data were acquired as part of the B-field Orion Protostellar Survey (BOPS, project ID: 2019.1.00086.S, PI: Ian Stephens) (*11*), with observations conducted on 29 November 2019. The data were taken using the compact ALMA configurations C43-1 and C43-2, as well as an intermediate configuration of both, providing baselines between approximately 15 and 314 m. This setup achieved an angular resolution of approximately 0.8" (corresponding to ~280 AU), with larger spatial scales being filtered out by the array. Observations were performed in full polarization mode (XX, YY, XY, YX) using the FDM. The spectral setup included four spectral windows: one centered on the CO (3–2) line at 345.796 GHz, with a velocity resolution of 0.423 km $s^{-1}$ (488 kHz channel spacing), and three continuum windows centered at 334.5, 336.5, and 348.5 GHz, each with a bandwidth of 1.875 GHz and a velocity resolution of ~0.87 km $s^{-1}$.

The standard calibration of the ALMA Band 6 and Band 7 visibilities was performed by ALMA staff as part of the Quality Assessment (QA2) process. We carried out additional phase-only self-calibration on the calibrated target data using CASA 5.4.0 (*29*, *30*), adopting adopting progressively shorter solution intervals down to the integration time. The final self-calibrated images show a substantial increase in signal-to-noise ratio (SNR), with improvements of approximately 1000 times for Band 6 and 4000 times for Band 7.

For continuum imaging, we first flagged bright spectral line channels in the self-calibrated data to prevent contamination of the continuum signal. Final continuum maps were produced using the tclean task in CASA, employing Briggs weighting with a robust parameter of 0.5. In Band 6, the final continuum image reaches an average noise level of ~2.4 mJy $beam^{-1}$, with a synthesized beam size of 1.60" × 0.84" (corresponding to ~ 460 AU) and a position angle (PA) of

-75°. In Band 7, the final Stokes $I$, $Q$, and $U$ maps have average noise levels of ~0.07, ~0.06, and ~0.06 mJy beam$^{-1}$, respectively. The synthesized beam for the Band 7 continuum image is 0.81" × 0.66" (corresponding to ~280 AU), with a PA of -80°. From the Stokes $Q$ and $U$ maps, we derived the debiased (*31*) polarized intensity $P$, polarization angle $\theta$, and fractional polarization $P_{\rm frac}$ using the standard definitions:

$$P = \sqrt{Q^2 + U^2 - \sigma^2}, \quad \theta = 0.5 \tan\left(U/Q\right), \quad P_{\rm frac} = P/I, \tag{S1}$$

where $\sigma$ is the noise level in the Stokes $Q$ or $U$ map.

For line imaging, we first performed continuum subtraction to remove residual continuum emission and enhance line sensitivity. The continuum-subtracted, self-calibrated data were then imaged using the tclean task in spectral cube mode. We used the $^{12}$CO (3–2) transition in Band 7 and the SiO (5–4) transition in Band 6 to trace the molecular outflow structures, while the C$^{17}$O (3–2) line in Band 7 and the C$^{18}$O (2–1) line in Band 6 were employed to probe the kinematics of the protostellar envelope.

## Accretion Streamer Model

To model the kinematics of the accretion streamer, we adopt the analytic ballistic infall solutions of Mendoza's model (*20*). In this mathematical framework, one considers an idealized spherical gas cloud of radius $r_0$, steadily accreting toward a central object located at the origin of a spherical coordinate system ($r$, $\theta$, $\phi$), where r denotes the radial distance, $\theta$ the polar angle, and $\phi$ the azimuthal angle. This construction is used solely to generate a family of ballistic trajectories; it is not intended to represent the actual morphology of the HOPS-182 envelope. All fluid particles located at $r_0$ are taken to follow rigid body motion in the azimuthal direction, i.e. they rotate about the $z$-axis of coordinates. At this position, particles also have a radial velocity component $v_{r0}$. The particle number density n at $r_0$ is taken to be constant with a value $n_0$. To simplify the problem, we neglect the self-gravity of the gas cloud, assuming that the gravitational potential is dominated by the central mass. In addition, we assume that pressure gradients are negligible compared to the kinetic and potential energies, allowing the particle trajectories to be approximated as ballistic. The total angular momentum of the gas cloud points in the direction of the z-axis, which contains the central object. In the limit where $r_0 \rightarrow \infty$ and $v_{r0} \rightarrow 0$, this model reduces to the classical ballistic accretion scenario described in earlier work (*32*, *33*).

Since the gas is accreted from $r_0$, its accretion flow rate per unit mass N at any fixed radial distance $r$ is constant and given by $N = 4\pi r_0^2 n_0 v_{r0} = const$. At the accretion radius $r_0$, the distribution of specific angular momentum $h$ is given by $h = h_0 \sin(\theta_0)$, where $h_0$ is the maximum specific angular momentum and $\theta_0$ is the initial polar angle of the particle's trajectory.

Under the assumptions outlined above, the specific energy $E$ and the specific angular momentum h are conserved along each particular trajectory. The total specific energy of particle $E$ is given by:

$$E = \frac{1}{2} v_r^2 + \frac{1}{2}\frac{h^2}{r^2} - \frac{GM}{r} = \frac{1}{2} v_{r0}^2 + \frac{1}{2}\frac{h_0^2 \sin^2\theta_0}{r_0^2} - \frac{GM}{r_0}, \tag{S2}$$

where $G$ is the gravitational constant, $M$ is the mass of the central object, and the subscript 0 denotes values at the cloud boundary $r_0$. Defining two dimensionless parameters $\mu$ and $\nu$, presented as the ratio of Ulrich's disk radius (*32*) to the original cloud's radius, and the initial amount of radial velocity measured in units of the Keplerian velocity (*34*):

$$\mu \equiv \frac{h_0^2}{r_0^2 E_0} = \frac{r_u^2}{r_0^2}, \quad \nu \equiv \frac{v_{r0}^2}{E_0}, \tag{S3}$$

where $r_{\rm u} = h_0^2/GM$ is the disk radius (*32*), and $E_0 \equiv GM/r_{\rm u}$ is the specific gravitational potential energy of a fluid particle evaluated at $r_{\rm u}$. With these parameters, the specific energy equation (Eq. S2) then becomes:

$$\varepsilon = \nu^2 + \mu^2 \sin^2 \theta_0 - 2\mu \,, \tag{S4}$$

where $\varepsilon \equiv 2E/E_0$ is the dimensionless energy.

Over the course of particle's motion, the trajectory was defined as a function of the parametric azimuthal angle $\varphi$ which at the initial position has the value $\varphi_0$. The trajectory of an infalling particle is defined by the solution to Kepler's problem (*34*):

$$r = \frac{\sin^2 \theta_0}{1 - e \cos \varphi} \,, \tag{S5}$$

where $e = (1+\varepsilon \sin 2\theta_0)^{1/2}$ is the eccentricity of the orbit. At the border of the cloud, $r = r_0 = 1/\mu$. Substituting this into Eq. S5 gives the following condition for $\varphi_0$:

$$\cos \varphi_0 = \frac{1}{e}\left(1 - \mu \sin^{-1} \theta_0\right) \,, \tag{S6}$$

Performing some spatial rotations, the following formulae between the angles $\varphi$, $\varphi^0$, $\theta$, $\theta^0$, $\phi$, and $\phi^0$ can be obtained:

$$\cos\left(\varphi - \varphi_0\right) = \frac{\cos\theta}{\cos\theta_0}, \quad \cos\left(\phi - \phi_0\right) = \frac{\tan\theta_0}{\tan\theta} \,. \tag{S7}$$

Using Eq. S7 to rewrite Eq. S5 yields:

$$r = \frac{\sin^2 \theta_0}{1 - e \cos \xi}, \quad \xi = \cos^{-1} \frac{\cos\theta}{\cos\theta_0} + \varphi_0 \,. \tag{S8}$$

In spherical coordinates, using the standard definitions of azimuthal $v_\phi = r\sin\theta \mathrm{d}\phi/\mathrm{d}t = h\sin\theta_0/(r\sin\theta)$, polar $v_\theta = r\mathrm{d}\theta/\mathrm{d}t = (\mathrm{d}\theta/\mathrm{d}\phi)(v\phi/\sin\theta)$, and radial $v_r = \mathrm{d}r/\mathrm{d}t = (\mathrm{d}r/\mathrm{d}\theta)(v_\theta/r)$ components of a velocity vector, the equations for velocity components are given by:

$$v_\phi = \frac{\sin^2\theta_0}{r\sin\theta}, \quad v_\theta = \frac{\sin\theta_0}{r\sin\theta}\left(\cos^2\theta_0 - \cos^2\theta\right)^{1/2}, \quad v_r = -\frac{e\sin\xi\sin\theta_0}{r(1 - e\cos\xi)} \,. \tag{S9}$$

We use this analytic ballistic model for test-particle trajectories in the potential of a point mass. The original derivation assumes an initially rotating spherical cloud to obtain a family of streamlines; in our application we use this solution purely as a kinematic template and fit a single trajectory to the position-position-velocity (PPV) structure of the observed streamer, without assuming that the actual envelope is spherical or that magnetic and pressure forces are dynamically negligible.

## TIPSY Fitting Procedure and Stellar Mass

TIPSY (Trajectory of Infalling Particles in Streamers around Young stars) (*19*) is a framework fitting theoretical trajectories expected for infalling gas, following Mendoza's model (*20*) to compute the positions (Eq. S5 to S7) and velocities (Eq. S9) of infalling particles. The fitting is done in three-dimensional (3D) PPV space: right ascension (RA), declination (Dec), and line-of-sight (LOS) velocity or radial velocity (RV). Following procedures previously described in (*19*), we fit the accretion streamer trajectory by:

(i) Compute the observed intensity-weighted means. As listed in Tab. S1, we visually choose a boundary for a sub-cube that encompasses the streamer emission using RA of -6.0" to -1.0", Dec of -10.0" to -1.0", and RV of 7.0 km s$^{-1}$ to 9.0 km s$^{-1}$ limits, with a threshold of specific noise level $\sigma = 4.0$ that used to eliminate pixels with low flux values. To get a more cleanly isolated streamer emission, we use the *sklearn* (*35*) implementation of the *OPTICS* (*36*) clustering algorithm, which computes density-based reachability distances to reveal clusters within a dataset. With the isolated and cleaned streamer emission, we first divide the streamer points into ten bins based on a distance metric of $d = (r^2+(r\theta)^2)^{1/2}$, where $r$ and $\theta$ are the polar coordinates of a point on the plane of the sky (POS) with respect to the protostar and the orientation of the streamer very close to the protostar, respectively (*19*). We then compute intensity-weighted means and intensity-weighted standard deviations of the RA, Dec, and RV values of all the points (red squares and lines in Fig. 2D).

(ii) Establish a parameter space. The space should cover all the possible initial conditions: relative position and velocity of the farthest point in 3D and the stellar mass. We determined the initial relative position as the physical distance to the protostellar system and the projected separation, and the relative velocity using the systemic velocity of the central protostar and the LOS velocity of the farthest point of the streamer to obtain the relative speed in the LOS direction. The LOS direction and POS velocities (the total velocity on the POS $(v_{RA}^2+v_{Dec}^2)^{1/2}$ and the initial direction of the particle on the POS $\arctan(v_{Dec}/v_{RA})$ are used instead of $v_{RA}$ and $v_{Dec}$ to reduce computations) are free parameters to be fitted.

(iii) Fit the trajectory of accretion streamer. Using the parameter space and the Mendoza's model (*20*), we calculate infalling trajectories for every parameter combination. The fitting trajectories are compared to the observed streamer curve (intensity-weighted means and standard deviations) to find the best fit. We independently compare the representative RA, Dec, and RV values using the distance metric, which is the independent variable for fitting. We use the first-order spline interpolation, as implemented in *scipy* (*37*), to get the theoretical values at the same distance metric values as the points of the observed streamer curve. Then we examine what fraction of the RA, Dec, and RV values of the observed streamer curve match within the error bars (standard deviations) to the theoretical values. This fraction is referred to as the "fitting fraction". We consider the best-fit trajectory as the one that can accommodate the highest fraction of mean values representing the observed streamer, within their error bars (as shown in yellow regions in the left panel of Fig. S1). In cases where multiple trajectories fit the same fraction of values, we choose the trajectory with the lowest chi-squared deviation as the best fit (as shown in yellow regions in the right panel of Fig. S1).

The stellar mass $M_*$ is an important parameter of the model, but there is currently no direct dynamical mass measurement for HOPS-182. We therefore use both the source luminosity and envelope properties to constrain a plausible range. HOPS-182 has a bolometric luminosity $L_{bol}$~71 $L_\odot$ and a bolometric temperature of ~51.9 K, indicative of a deeply embedded and actively accreting system (*15*, *17*). Radiative transfer modeling of its spectral energy distribution (SED) yields an envelope mass of ~0.33 $M_\odot$ and an envelope mass infall rate $\dot{M}_{env}$ = 2.5 × 10$^{-6}$ $M_\odot$ year$^{-1}$ (*18*). For a deeply embedded Class 0 source, since $L_{bol} = L_* + L_{acc}$, where $L_*$ and $L_{acc}$ are the stellar and accretion luminosity, respectively, then we get:

$$L_{bol} > L_{acc} = GM_*\dot{M}_*/R_* \,, \quad \text{(S10)}$$

where $R_*$ is the protostellar radius, and $\dot{M}_*$ is the stellar accretion rate. If we assume the stellar accretion rate is comparable to the envelope infall rate $\dot{M}_* \approx \dot{M}_{env}$ and adopt a typical low/intermediate-mass protostellar radius $R_* \approx 3\ R_\odot$, we obtain a stellar mass $M_* < 2.5\ M_\odot$. Besides, stellar masses below 0.8 $M_\odot$ do not reproduce the observed $C^{18}O$ streamer kinematics with TIPSY. To be conservative, we adopt $0.8 \le M_* \le 2.0\ M_\odot$ as the allowed mass range, which brackets the values expected for HOPS-182 with similar luminosity, envelope mass, and accretion rate. Within this mass range, the projected streamer trajectory in RA-Dec and the velocity locus in the PV diagram remains essentially unchanged (Fig. S2). The identification of the structure as a gravitationally bound infall streamer is therefore not sensitive to the exact value of $M_*$. The main effect of varying $M_*$ is to rescale of the dimensional velocities, specific angular momentum, and infall timescales. Tab. S2 lists the best-fit parameters for each value of $M_*$. The infall timescale $T_{inf}$, defined as the time taken for the best-fit solutions to travel from the farthest point in the observed streamer to the point closest to the protostars, ranges from (5.1 ± 1.4) × 10$^4$ for $M_*$ = 0.8 $M_\odot$ to (1.7 ± 0.1) × 10$^4$ years for $M_*$ = 2.0 $M_\odot$.

**Streamers-Magnetic-Field Alignment**

To quantify the misalignment between the observed and the best-fit TIPSY streamer trajectories and the local magnetic field orientation, we measure, at each location, the angle between the magnetic field orientation and the tangent to the streamers. For both trajectories, we compute the cumulative arc length along these points and apply cubic-spline interpolation to obtain a smooth curve in the image plane. The local streamer direction at each sampled position is then given by the tangent to this spline, i.e. by the finite-difference derivative of the interpolated trajectory. To define a magnetic field trajectory that is directly comparable to this model streamer, we construct a magnetic field streamline by integrating along the magnetic field orientation map. This streamline is started at the outermost point of the streamer (farthest from the source center), and is then followed inward along the local magnetic field. The magnetic field trajectory tracing is stopped when the distance from a point along the field line to the protostar equals the distance from the innermost point of the streamers (closest to the source center) to the protostar, so that the streamers and the magnetic field trajectory cover the same radial range.

Using Nyquist Sampling into account, we identify the nearest points on both the TIPSY trajectory and the magnetic field streamline, and compute the signed angle differences $\Delta\phi$, which we wrap into the interval [-90°, 90°]. From the starting point of the TIPSY fit, we found that in the first ~1000 AU, the magnetic field follows the observed streamer trajectory better than the TIPSY one. The mean orientation angle difference between the streamer and the magnetic field trajectory is ~2°, and between the streamer and the TIPSY is ~10°. Closer to the protostars, the observed streamer and the magnetic field begin deviate from one to another (mean angle difference ~32°), while the streamer lies somewhat closer to the TIPSY trajectory (mean difference ~26°).

## Physical Properties of the Streamer

We estimate the gas mass of the streamer using the 0.87 mm dust continuum emission under the assumption of optically thin thermal emission. The gas mass Mgas is computed as:

$$M_{\mathrm{gas}} = \eta M_{\mathrm{dust}} = \eta \frac{S_\nu D^2}{\kappa_\nu B_\nu(T)}, \tag{S11}$$

where $\eta$ = 100 is the gas-to-dust mass ratio, $D$ = 390 pc is the distance to HOPS-182, $S_\nu$ is the total flux density, $\kappa_\nu \approx 1.84$ cm$^2$ g$^{-1}$ is the dust opacity at 0.87 mm wavelength (*38*), $\nu$ is the frequency corresponding to the observed wavelength, and $B_\nu(T)$ is the Planck function evaluated at dust temperature of $T$, expressed as:

$$B_\nu(T) = \frac{2h\nu^3}{c^2}\left(\frac{1}{e^{h\nu/k_B T}} - 1\right), \tag{S12}$$

with $h$ as Planck's constant, $c$ the light speed, and $k_B$ the Boltzmann's constant. In the case of a dusty cloud heated by internal stellar radiation, the radial temperature profile around an Orion protostar can be expressed as (*15*):

$$T(r) = T_0\left(\frac{L_{\mathrm{bol}}}{L_\odot}\right)^{0.25}\left(\frac{r}{50\ \mathrm{au}}\right)^{\beta}, \tag{S13}$$

where $T_0$ = 43 K is the average temperature for a ~1 $L_\odot$ protostar at a radius of ~50 AU, $L_{\mathrm{bol}}$ = 71 $L_\odot$ is the bolometric luminosity, $r$ is the distance from the protostar, and $\beta$ is the temperature power-law index. Previous work suggest $\beta \gtrsim -0.4$ for Class 0 protostars (*39*), and $\beta \sim$ -0.46 to -0.50 for Orion protostellar systems (*15*, *40*). However, this profile shows a slight decline in the outer 100 AU regions due to back-heating from the surrounding envelope, so for HOPS-182 we adopt a somewhat shallower index $\beta \sim -0.4$. The clear detection of $C^{17}O$ and $C^{18}O$ emission in HOPS-182 implies that dust temperature cannot fall below 20 K, since CO would otherwise be largely frozen out and undetectable (*41*). Excluding the extremely high temperature in the central pixel, we obtain an average dust temperature of ~43.7 K at a radius of 1200 AU, which covers a detection of continuum intensity at least $10\sigma$.

The dominant sources of uncertainty are dust temperature $T$, dust opacity $\kappa_\nu$, and flux calibration. We explicitly include a conservative 30% uncertainty for potential temperature variations based on the considerations above. For the dust opacity, we use the standard value from the literature (*38*), and considering a factor of 2 in opacity. The absolute flux calibration at 0.87 mm is typically accurate to within ~10%, which is also included in our final mass uncertainty budget. All combined, we derived uncertainties of about ~ 40% for the masses and densities using standard error-propagation techniques (*26*).

To estimate the streamer density, we assume a cylindrical morphology, justified by the filament's aspect ratio and centrally directed gravitational acceleration. The density is given by:

$$\rho_{\text{str}} = \frac{M_{\text{gas}}}{V_{\text{str}}} = \frac{M_{\text{gas}}}{\pi r_{\text{str}}^2 L_{\text{str}}}, \quad \text{(S14)}$$

where $r_{\text{str}}$ is the streamer's radius of the cross-section and $L_{\text{str}} = 1.7 \times 10^3$ AU is the path length of the best-fit trajectory. To derive $r_{\text{str}}$, we extract transverse cuts of the 0.87 mm continuum intensity perpendicular to the streamer's axis. Gaussian fitting of these profiles (excluding edge regions) yields a median full width at half maximum (FWHM) of 627 AU. After deconvolving the synthesized beam size of 280 AU, we adopt half of the deconvolved FWHM as the streamer radius, yielding $r_{\text{str}} = 281$ AU. For the best-fit streamer structure, the total flux density is $S_\nu = 0.23$ Jy, yielding a streamer mass of $M_{\text{str}} \approx 0.07\ M_\odot$ (derived from Eq. S11), then the streamer density is $\rho_{\text{str}} = 9.5 \times 10^{-17}$ g cm$^{-3}$. Assuming a 40% uncertainty on the streamer mass, the corresponding mass infall rate of the streamer $M_{\text{str}}/T_{\text{inf}}$ ranges from $(1.2 \pm 0.6) \times 10^{-6}$ to $(3.9 \pm 1.6) \times 10^{-6}\ M_\odot$ year$^{-1}$ for a range of stellar mass from 0.8 to 2.0 $M_\odot$.

## Magnetic Properties of the Streamer

To estimate the plane-of-sky magnetic field strength, we apply the Davis-Chandrasekhar-Fermi (DCF) method (*42*, *43*):

$$B \approx Q \frac{\sqrt{4\pi \rho_{\text{str}}}}{\delta\phi_{\text{str}}} \sigma_{\text{nth}}, \quad \text{(S15)}$$

where $\delta\phi_{\text{str}} = 21.6° \pm 3.1°$ is the measured dispersion of polarization angles in the streamer, and Q ≈ 0.5 is a correction factor appropriate for small angular dispersions ($\delta\phi_{\text{str}} < 25°$) (*44*). The non-thermal velocity dispersion $\sigma_{\text{nth}}$ is calculated as:

$$\sigma_{\text{nth}} = \sqrt{\sigma_{\text{v}}^2 - \sigma_{\text{th}}^2}, \quad \text{(S16)}$$

where $\sigma_{\text{th}} = (k_{\text{B}}T/m_{\text{obs}})^{1/2}$ is the thermal velocity dispersion. We adopt a kinetic temperature 43.7 K with a 30% uncertainty, and a molecular mass of $m_{\text{obs}} = 30m_{\text{H}}$ (where $m_{\text{H}}$ is the mass of the hydrogen atom) for $C^{18}O$. This yields $\sigma_{\text{th}} = 0.110 \pm 0.001$ km s$^{-1}$.

The observed line-of-sight velocity dispersion $\sigma_{\text{obs}}$ is corrected for spectral resolution to obtain the intrinsic velocity dispersion $\sigma_{\text{v}}$:

$$\sigma_{\text{v}} = \sigma_{\text{obs}}^2 - \sigma_{\text{res}}^2 = \sigma_{\text{obs}}^2 - \frac{\Delta v_{\text{r}}^2}{8\ln 2}, \quad \text{(S17)}$$

where $\Delta v_{\text{r}} \approx 0.04$ km s$^{-1}$ is the channel width of the $C^{18}O$ observations. From the moment 2 map of the $C^{18}O$ (2–1) line, we measure an average $\sigma_{\text{obs}}$ across the streamer and derive a corrected value of $\sigma_{\text{v}} = 0.33 \pm 0.01$ km s$^{-1}$, yielding a non-thermal component $\sigma_{\text{nth}} = 0.31 \pm 0.01$ km s$^{-1}$. Using a mean density $\rho_{\text{str}} = 6.9 \times 10^{-17}$ g cm$^{-3}$ (with a 40% uncertainty) and applying Eq. S15, we estimate the plane-of-sky magnetic field strength of the streamer to be $\boldsymbol{B} = (1.4 \pm 0.5)$ mG. However, the DCF method used to estimate the magnetic field strength is know to carry systematic uncertainties of order a factor of ~ 2 (*25–27*), due to departures from the ideal DCF assumptions (e.g., homogeneity, purely Alfvénic fluctuations, and small-angle perturbations), we therefore adopt this uncertainty for the following estimates.

## Magnetic Tension and Gravitational Force

To quantify the magnetic tension force along the streamer, we follow the formula (*22*, *23*):

$$f_{\text{mag}} = \frac{B^2 \kappa}{4\pi}, \tag{S18}$$

where $\kappa$ is the curvature of the magnetic field line, measured along the field orientation from the initial point of the best-fit streamer (farthest from the source center), to the position when the distance from a point along the field line to the protostar equals the distance from the innermost TIPSY point (closest to the source center) to the protostar. We estimate $\kappa$ by fitting a best-fit circle to the entire trajectory. The fitting was done using a geometric least-squares approach, solving for the circle center and radius ($x_c$, $y_c$, $R_c$) that minimize the residuals between the observed points and the circular arc. The corresponding global curvature is defined as $\kappa = 1/R_c$. To assess the uncertainty in the curvature measurement, we performed 1000 bootstrap realizations of the field line by resampling the trajectory points with replacement and repeating the circle-fitting procedure for each realization. Then the standard deviation of the curvature distribution were adopted as the uncertainty of the estimated curvature. As shown in Fig. S3, the fitted radius is $R_c = (2.2 \pm 0.2) \times 10^3$ AU, thus the curvature $\kappa = (3.1 \pm 0.2) \times 10^{-17}$ cm$^{-1}$. For a magnetic field strength of the streamer $B$ = 1.4 mG with a factor-of-2 uncertainty, the magnetic tension is estimated to be $f_{\text{mag}} = (0.5\text{–}2.1) \times 10^{-23}$ dyn cm$^{-3}$. Removing the ordered component magnetic field of the trajectory, we fit the curvature to be $\kappa = (5.0 \pm 0.3) \times 10^{-17}$ cm$^{-1}$ (see Fig. S3). This yields a maximum magnetic tension to be $f_{\text{mag,max}} = (0.8\text{–}3.3) \times 10^{-23}$ dyn cm$^{-3}$.

The gravitational force is given by (*22*, *23*):

$$f_{\text{grav}} = \frac{G M_* \rho_{\text{str}}}{D_{\text{str}}^2}, \tag{S19}$$

where $G$ is the gravitational constant, $M_*$ is the stellar mass, $\rho_{\text{str}} = 9.5 \times 10^{-17}$ g cm$^{-3}$ is the streamer density (with 40% uncertainty), and $D_{\text{str}} = (\text{RA}^2+\text{Dec}^2)^{1/2} = (2.2 \pm 0.2) \times 10^3$ AU is the intensity-weighted mean distance to the protostar (see Tab. S2). Using Eq. S19, the gravitational force $f_{\text{grav}} = (0.5\text{–}3.1) \times 10^{-23}$ dyn cm$^{-3}$ across the allowed stellar-mass range. At the position where the magnetic field curvature reaches its maximum, with the mean distance $D_{\text{str}} = (1.3 \pm 0.1) \times 10^3$ AU, the corresponding gravitational force $f_{\text{grav,max}} = (2.0\text{–}9.4) \times 10^{-23}$ dyn cm$^{-3}$.

## Angular-velocity-to-magnetic-field-strength Ratio

To determine whether the dynamics is dominated by rotation or magnetic regulation, we can estimate the observed ratio of angular velocity to magnetic field strength $\omega/B$. If the ratio is larger than the critical value, the centrifugal forces dominate the dynamics, otherwise the magnetic forces regulate the dynamics. The critical value is expressed as (*24*):

$$\left(\frac{\omega}{B}\right)_{\text{crit}} = 1.69 \times 10^{-7} \times \left(\frac{c_s}{0.19 \text{ km s}^{-1}}\right)^{-1}, \tag{S20}$$

where $c_s = (k_B T/\mu_{\text{gas}} m_H)^{1/2}$ is the sound speed, with $\mu_{\text{gas}} = 2.33$ the mean molecular weight (*45*). For an envelope gas temperature of $T$ = 43.7 K with 30% uncertainty, the sound speed is $c_s = 0.40 \pm 0.04$ km s$^{-1}$. Thus the critical value is $(\omega/B)_{\text{crit}} = (8.2 \pm 0.8) \times 10^{-8}$ year$^{-1}$ μG$^{-1}$.

For the TIPSY best-fit streamer, the specific angular momentum $j$ ranges from $(3.3 \pm 1.1) \times 10^2$ to $(5.8 \pm 0.7) \times 10^2$ AU km s$^{-1}$ (see Tab. S2), thus the angular velocity $\omega = j/D_{\text{str}}^2$ ranges from $(1.4 \pm 0.5) \times 10^{-5}$ to $(2.5 \pm 0.5) \times 10^{-5}$ year$^{-1}$. With the magnetic field strength of the streamer $B$ = 1.4 mG, the observed ratio of angular velocity to magnetic field strength $\omega/B = (0.5–3.6) \times 10^{-8}$ year$^{-1}$ $\mu$G$^{-1}$, which is more than two times smaller than the critical value.

## Gravitational Field

To compute the gravitational field across the envelope, we first estimated the gas mass in each pixel of the Stokes $I$ map using Eq. S11, assuming optically thin dust emission. The dust temperature in each pixel was calculated using Eq. S13. Using the derived mass map, and taking into account the stellar mass at the source center, the gravitational potential was computed by summing the contributions from all other pixels, each weighted by its mass and its inverse distance from the pixel of interest (*21*). The local gravitational field vector at each position was then calculated as the gradient of this potential, yielding both the strength and orientation of the gravitational acceleration across the map (see Fig. S4). Note that the gravitational field patterns are essentially unchanged over the adopted stellar-mass range.

## Uncertainties

Unless otherwise noted, quoted errors in the main text correspond to statistical uncertainties propagated from measurement errors. In this work, several parameters are obtained directly from fitting and analysis of the data, such as the length and 3D geometry of the best-fit streamer, the specific angular momentum, and the infall timescale from the TIPSY modeling. Their quoted uncertainties reflect the scatter of acceptable solutions and the formal fitting errors (e.g., through the fitting fraction and the spread of trajectories consistent with the observational error bars). On the other hand, a separate class of uncertainties arises from physical assumptions that enter the conversion from observables to physical parameters. These include the envelope temperature structure, the dust-based streamer mass, and the magnetic field curvature. These uncertainties are propagated into derived quantities using standard error-propagation techniques. In addition, the DCF method used to estimate the magnetic field strength is know to carry systematic uncertainties of order a factor of ~2 (*24–26*), due to departures from the ideal DCF assumptions (e.g., homogeneity, purely Alfvénic fluctuations, and small-angle perturbations).

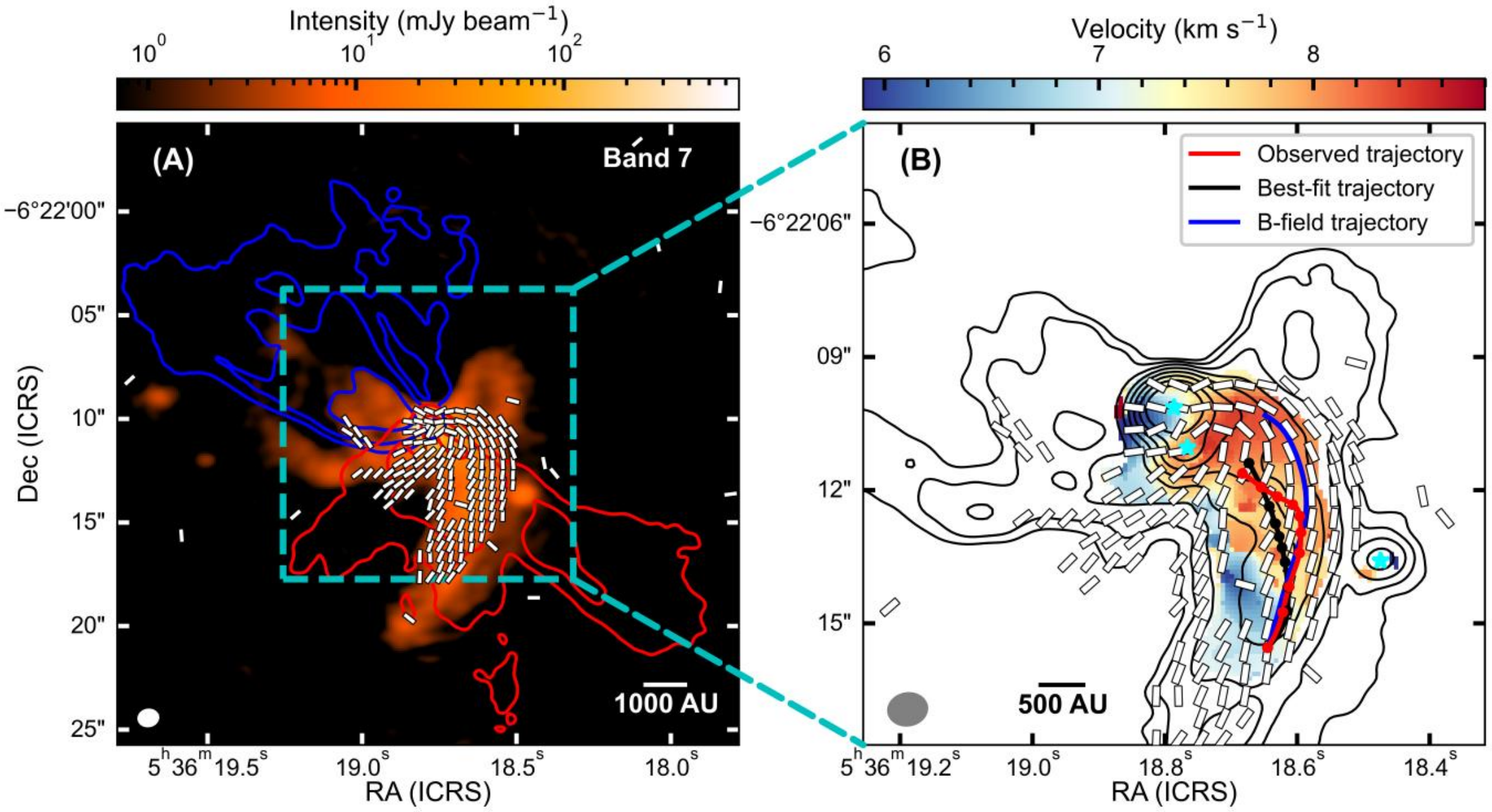

**Fig. 1: ALMA Band 7 (345 GHz, 0.87 mm) observations. (A):** Dust emission shown in color scale, overlaid with red and blue contours showing the red- and blue-shifted CO (3–2) emission from the bipolar molecular outflow, and white segments indicating magnetic field orientation. Outflow contour levels are set at 5 times the rms (70 mJy beam$^{-1}$) × (1, 10, 50). **(B):** Zoom-in view (14", ~5400 AU) of panel A showing the velocity fields traced by $C^{17}O$ (3–2) in color scale, overlaid with the dust intensity contours and magnetic field segments in white. Red points and curve indicate the intensity-weighted means and observed streamer trajectory, while black points and curve show the TIPSY best-fit solutions and trajectory. Blue curve indicates the magnetic field trajectory. Cyan stars highlight the location of protostars. Intensity contour levels are set at 10 times the rms (0.07 mJy beam$^{-1}$) × (1, 2, 4, 8, 16, 32, 64, 128, 256).

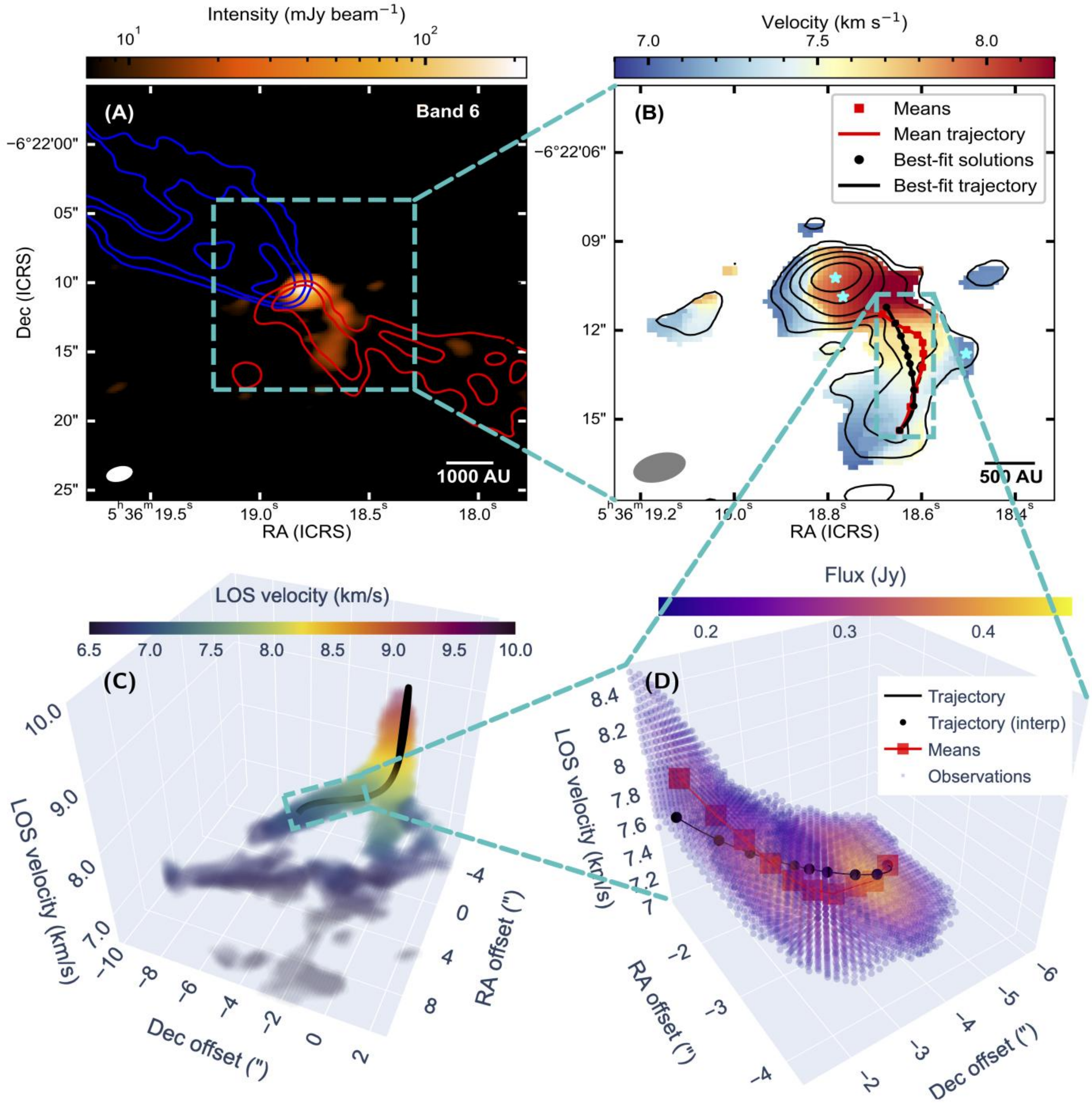

**Fig. 2: ALMA Band6 (220 GHz, 1.36 mm) observations and the best fit for the accretion streamer model using the TIPSY framework. (A):** 1.36 mm dust emission shown in color scale, overlaid with contours showing the red- and blue-shifted SiO (5–4) emission from the bipolar outflow. Outflow contour levels are set at 5 times the rms (0.45 Jy beam$^{-1}$) × (1, 2, 4). **(B):** A zoom-in view (14", ~5400 AU) of panel A showing the velocity fields traced by $C^{18}O$ (2–1) in color scale, overlaid with black contours tracing the dust emission. Black points and curve indicate the best-fit solutions and the TIPSY best-fit trajectory of the accretion streamer, while red points and curve indicate intensity-weighted means and its trajectory. These solutions and trajectories are projected onto the plane of the sky from PPV space. Cyan stars highlight the location of protostars. Intensity contour levels are set at 10 times the rms (2.4 mJy beam$^{-1}$) × (1, 2, 4, 8, 16). **(C):** Isometric projection of the three-dimensional position-position-velocity (PPV) diagram of pixels with $C^{18}O$ intensities $>5\sigma$ in the whole field of view, overlaid with the best-fit trajectory represented by the black curve. **(D):** Isometric projection of the PPV diagram of an isolated and cleaned streamer. The red square represents the dust intensity-weighted mean position in the PPV diagram, while the black circles denote the best-fit solutions for the means, and the black curve indicates the best-fit trajectory.

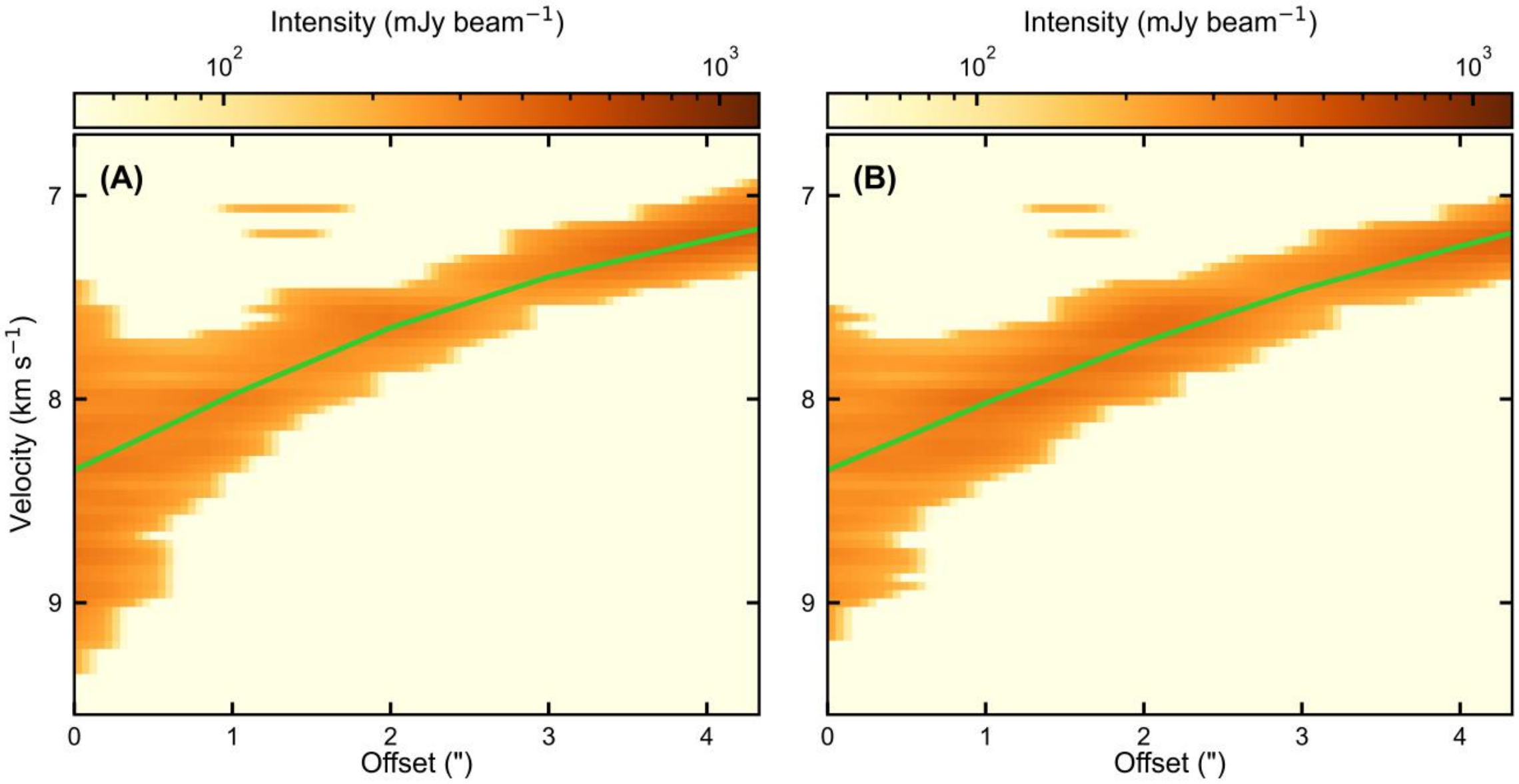

**Fig. 3: Position-velocity diagrams along the best-fit trajectory (A) and the observed streamer spine (B).** Color scale indicates the $C^{18}O$ (2–1) intensity. Offset = 0 is the point of the streamer closest to the protostar. Green lines highlight the paths along the highest intensity.

**Acknowledgments:** We thank the ALMA staff for performing the observations and quality assessment of the data.

**Funding:** BH, JMG, and AS-M acknowledge support by grant PID2023-146675NB-I00 (MCI-AEI-FEDER, UE). BH acknowledges support by the Korea Astronomy and Space Science Institute under the R&D program (Project No. 2026-1-844-00) supervised by the Ministry of Science and ICT. AS-M acknowledges support from the RyC2021-032892-I grant funded by MCIN/AEI/10.13039/501100011033 and by the European Union 'Next GenerationEU'/PRTR. This work is also partially supported by the program Unidad de Excelencia María de Maeztu CEX2020-001058-M. LWL acknowledges support from NSF AST-1910364 and NSF AST-2307844. JALV acknowledge support from the National Science and Technology Council (NSTC) of Taiwan through grant 112-2112-M-001-039-MY3. PS was partially supported by the JSPS Grant-in-Aid for Scientific Research KAKENHI Number JP23H01221. VJMLG acknowledges supports by the Spanish program Unidad de Excelencia María de Maeztu CEX2020-001058-M, financed by MCIN/AEI/10.13039/501100011033, and by the MaX-CSIC Excellence Award MaX4-SOMMA-ICE, and by the European Research Council (ERC) under the European Union's Horizon 2020 research and innovation program (grant agreement No. 101098309 - PEBBLES). EL acknowledges support by the Guangxi Science and Technology Innovation Platform Program (Leitai Action Plan, grant No. Guike LT2600640026), by the Guangxi Key R&D Program (Guangxi Funeng Action Plan, grant No. Guike FN2504240040), and by the Guangxi Talent Program (Highland of Innovation Talents).

**Author contributions:**
Conceptualization: BH, JMG, IWS, TM, VJMLG, PC, MFL, LWL, ASM, SS, QZ, CLHH, NK, EL
Methodology: BH, JMG, IWS, TM, VJMLG, NMM, MFL, LWL, JALV, JEP, ASM, QZ, CLHH
Software: BH, VJMLG, JALV, JEP, CLHH
Validation: BH, JMG, VJMLG, PC, MFL, JALV, PS, QZ, EL
Formal analysis: BH, JMG, TM, VJMLG, PC, MFL, JALV, JEP, SS, CLHH
Investigation: BH, JMG, TM, VJMLG, PC, JEP, CLHH
Resources: JMG, IWS
Data curation: BH, PC
Writing–original draft: BH, VJMLG
Writing–review & editing: BH, JMG, IWS, TM, VJMLG, NMM, PC, MFL, ZYL, LWL, JALV, JEP, ASM, PS, SS, QZ, NK, EL, PCM
Visualization: BH, VJMLG, JEP
Supervision: JMG, VJMLG, CLHH
Project administration: BH, JMG, CLHH
Funding acquisition: JMG, ASM, CLHH

**Competing interests:** The authors declare they have no competing interest.

**Data, code, and materials availability:** This paper makes use of the following ALMA data: ADS/JAO.ALMA#2019.1.00086.S and ADS/JAO.ALMA#2018.1.01073.S. ALMA is a partnership of ESO (representing its member states), NSF (USA) and NINS (Japan), together with NRC (Canada), MOST and ASIAA (Taiwan), and KASI (Republic of Korea), in cooperation with the Republic of Chile. The Joint ALMA Observatory is operated by ESO, AUI/NRAO and NAOJ. The National Radio Astronomy Observatory is a facility of the National Science Foundation operated under cooperative agreement by Associated Universities, Inc. The original observational data, along with the associated calibration scripts, are available from the ALMA Science Archive at https://almascience.nrao.edu/aq/ using the program codes above. The final fits files, models, and scripts necessary to reproduce the data products, results, and figures presented in this work are publicly available at Harvard Dataverse: https://doi.org/10.7910/DVN/9HVOZN. The original TIPSY model for streamer fitting is available at https://github.com/AashishGpta/TIPSY. PV diagrams were generated using pvextractor, available at https://pvextractor.readthedocs.io/en/latest/index.html. This work also made use of Astropy: a community-developed core Python package and ecosystem of tools and resources for astronomy (www.astropy.org).

## Supplementary Materials

This PDF file includes:
Figures S1–S4
Tables S1–S2

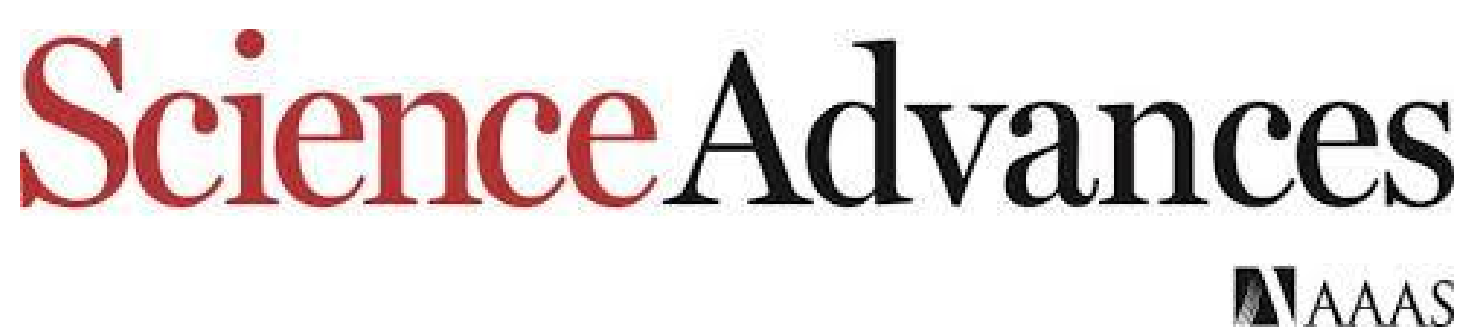

# Supplementary Materials for

## Magneto-Gravitational Regulated Streamer Accretion onto a Class 0 Protostellar System

Bo Huang *et al.*

*Corresponding author: Bo Huang, huang@ice.csic.es

**This PDF file includes:**

Figs. S1 to S4
Tables S1 to S2

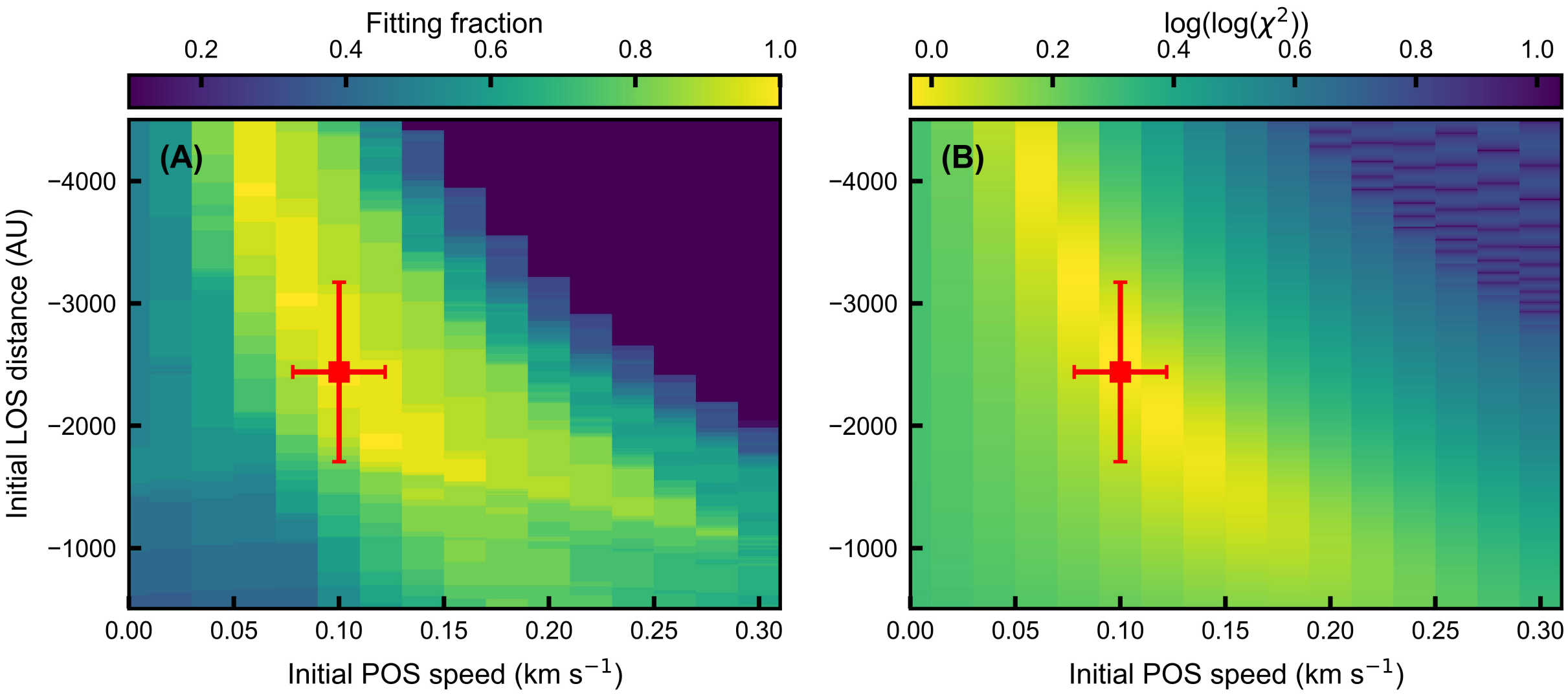

**Fig. S1: Distribution of goodness-of-fit estimates as functions of free parameters: initial speed on the POS (x-axis) and initial spatial offset in the LOS direction (y-axis). (A):** distribution of fractions of coordinate values of points in the observed streamer curve (intensity-weighted means and standard deviations), which is consistent with the theoretical trajectories. **(B):** distribution of log(log($\chi^2$)) deviations between the observed streamer curve and theoretical trajectories. In all the plots, yellow regions represent good fits. Red squares represent the best fit, while the red lines passing through them represent errors.

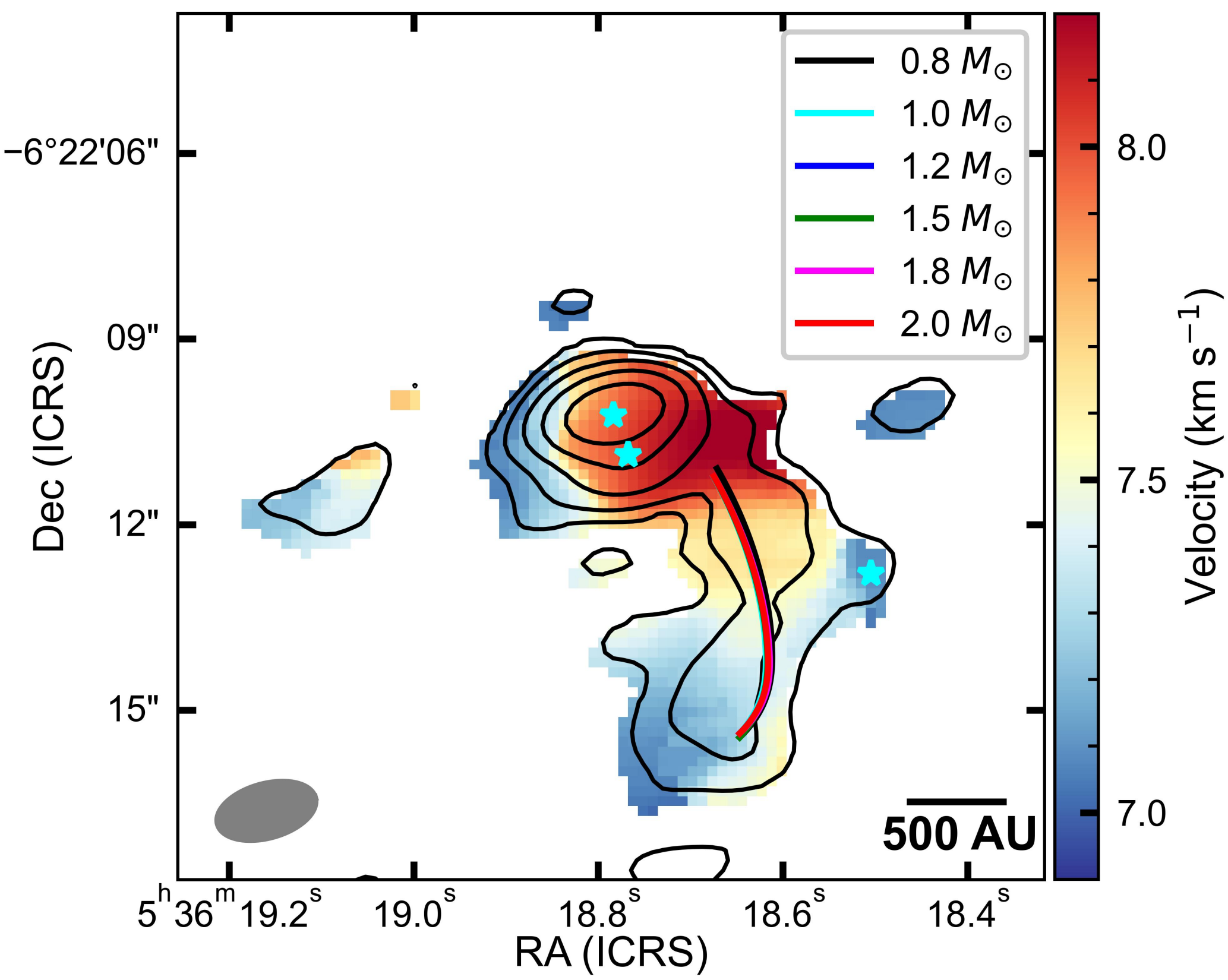

**Fig. S2: The TIPSY best-fit trajectories for different stellar masses.** The color scale is the velocity field derived from $C^{18}O$ (2–1), overlaid with black contours of the 1.36 mm dust continuum emission. Continuum contours are plotted at (1, 2, 4, 8, 16) × 10$\sigma$, where $\sigma$ = 2.4 mJy beam$^{-1}$. Cyan stars highlight the location of protostars.

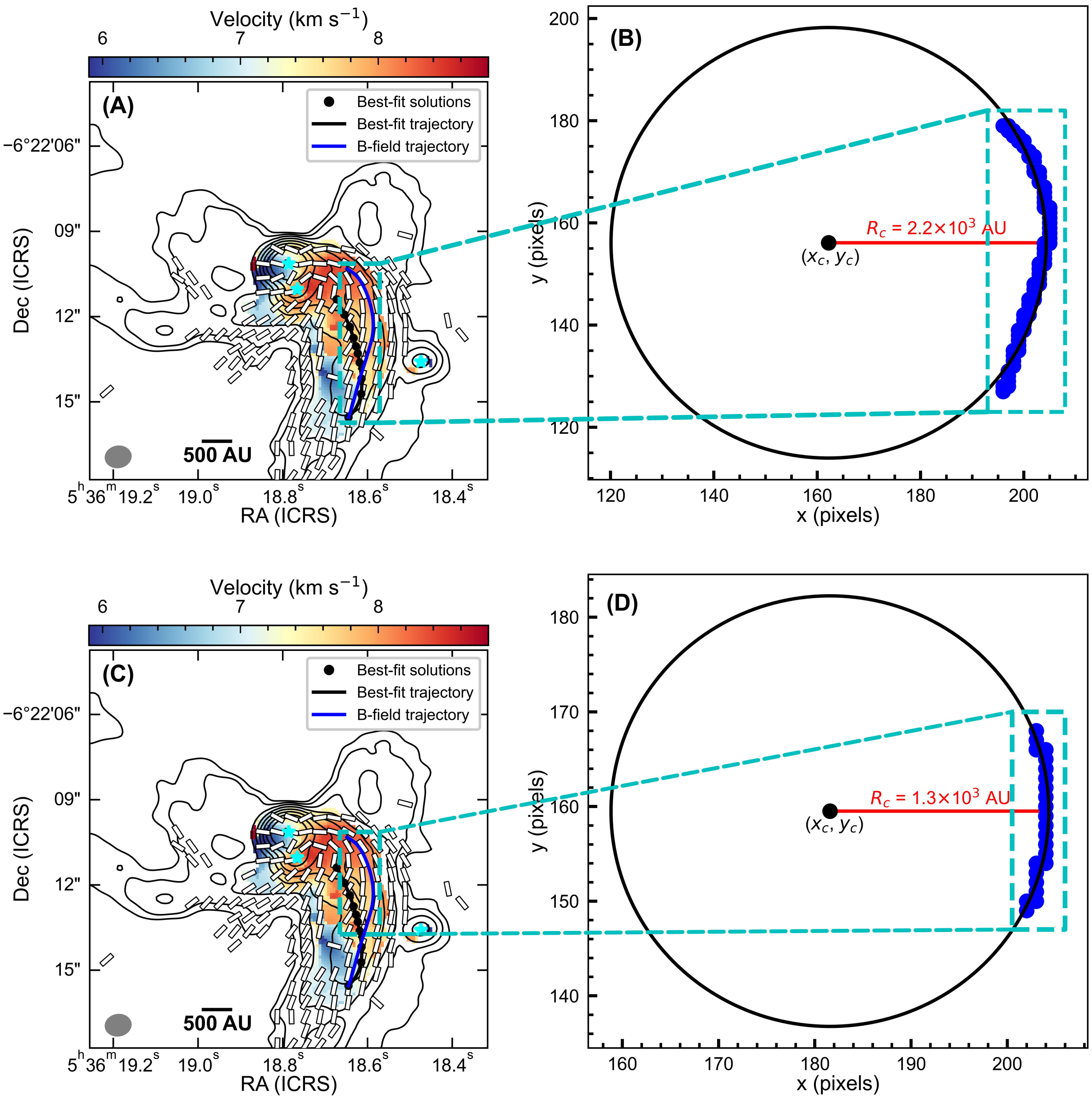

**Fig. S3: Fitted curvature of the magnetic field line. Left panels (A and C):** Velocity fields traced by $C^{17}O$ (3–2) are shown in color scale, overlaid with 0.87 mm dust continuum intensity contours in black and magnetic field segments in white. Black points and curve indicate the TIPSY best-fit positions and trajectory of the accretion streamer, while the blue curve highlights the selected magnetic field line. Cyan stars highlight the location of protostars. Intensity contour levels are set at 10 times the rms (0.07 mJy beam$^{-1}$) × (1, 2, 4, 8, 16, 32, 64, 128, 256). **Right panels (B and D):** Illustration of the magnetic field curvature fitting. Magenta points trace the selected magnetic field line, and the black circles denote the best-fit circles to the field line. In the bottom panels, the ordered component of the magnetic field has been subtracted to highlight the local curvature.

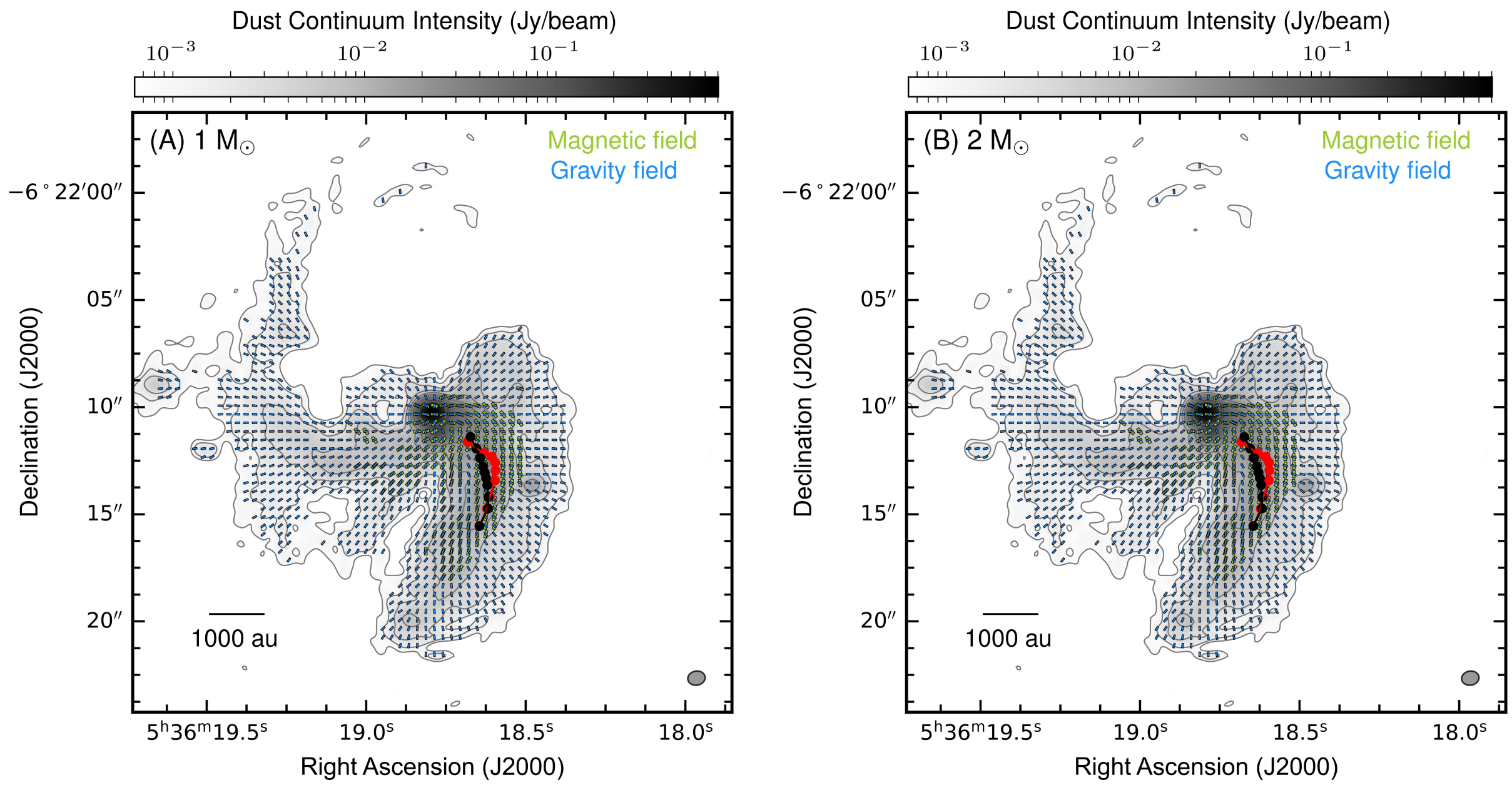

**Fig. S4: Gravitational field diagrams for stellar mass of 1 $M_{\odot}$ (left) and 2 $M_{\odot}$ (right).** 0.87 mm dust intensity map in gray scale, overlaid with gravitational field segments in blue and magnetic field segments in green. Gray contours indicate the 0.87 mm dust continuum intensity, with contour levels set at 10 times the rms (0.07 mJy beam$^{-1}$) × (1, 2, 4, 8, 16, 32, 64, 128, 256). The best-fit trajectory and the intensity-weighted means are show with the black and red lines, respectively.

| Parameters | HOPS-182 |
|---|---|
| Stellar mass ($M_\odot$) | 0.8/1.0/1.2/1.5/1.8/2.0 |
| Distance (pc) | 390 |
| Systemic velocity (km $s^{-1}$) | 7.1 |
| Min. RV offset (km $s^{-1}$) | 7.0 |
| Max. RV offset (km $s^{-1}$) | 9.0 |
| Min. RA offset (") | -6.0 |
| Max. RA offset (") | -1.0 |
| Min. Dec offset (") | -10.0 |
| Max. Dec offset (") | -1.0 |
| Significance level ($\sigma$) | 4.0 |

**Tab. S1: Parameters used to isolate and fit the HOPS-182 streamer.** Rows 2 to 11 list the stellar mass, source distance, systemic velocity, minimum and maximum radial velocity (RV) offsets, minimum and maximum right ascension (RA) offsets, minimum and maximum declination (Dec) offsets, and the threshold of the specific noise level, respectively. Offsets are relative to the source center.

| Stellar mass ($M_\odot$) | 0.8 | 1.0 | 1.2 | 1.5 | 1.8 | 2.0 |
|---|---|---|---|---|---|---|
| RA offset (AU) | -849±269 | -849±269 | -849±269 | -849±269 | -849±269 | -849±269 |
| Dec offset (AU) | -2044±132 | -2044±132 | -2044±132 | -2044±132 | -2044±132 | -2044±132 |
| LOS offset (AU) | -2960 | -2440±734 | -2210±966 | -1740±274 | -1340±185 | -1240±152 |
| RA speed (km $s^{-1}$) | -0.07±0.02 | -0.09±0.02 | -0.11±0.03 | -0.14±0.02 | -0.19±0.02 | -0.20±0.02 |
| Dec speed (km $s^{-1}$) | 0.03±0.01 | 0.04±0.01 | 0.04±0.02 | 0.07±0.02 | 0.07±0.02 | 0.10±0.02 |
| LOS speed (km $s^{-1}$) | -0.06±0.07 | -0.06±0.07 | -0.06±0.07 | -0.06±0.07 | -0.06±0.07 | -0.06±0.07 |
| Kinetic energy ($10^{-3}$ $km^2$ $s^{-2}$) | 5.2±4.6 | 7.0±4.6 | 9.2±5.4 | 14.8±5.1 | 22.0±5.7 | 26.2±6.2 |
| Potential energy ($km^2$ $s^{-2}$) | -0.19±0.04 | -0.27±0.05 | -0.34±0.08 | -0.47±0.04 | -0.62±0.04 | -0.70±0.04 |
| Angular momentum (AU km $s^{-1}$) | 325±110 | 356±94 | 399±121 | 476±76 | 535±70 | 579±74 |
| Infall time ($10^4$ year) | 5.11±1.42 | 3.76±1.09 | 3.15±1.30 | 2.35±0.20 | 1.90±0.10 | 1.72±0.07 |
| $f_{grav}$ ($10^{-24}$ dyn $cm^{-3}$) | 3.0±1.3 | 3.8±1.6 | 4.5±1.9 | 5.7±2.4 | 6.8±2.8 | 7.5±3.2 |
| $f_{grav,max}$ ($10^{-23}$ dyn $cm^{-3}$) | 1.7±0.7 | 2.1±0.9 | 2.5±1.1 | 3.1±1.3 | 3.8±1.6 | 4.2±1.8 |
| $\omega$ ($10^{-5}$ $yr^{-1}$) | 1.4±0.5 | 1.5±0.5 | 1.7±0.6 | 2.0±0.4 | 2.3±0.4 | 2.5±0.5 |
| $\omega/B$ ($10^{-8}$ $yr^{-1}$ $\mu G^{-1}$) | 0.5 – 2.0 | 0.6 – 2.2 | 0.6 – 2.5 | 0.7 – 3.0 | 0.8 – 3.3 | 0.9 – 3.6 |

**Tab. S2: Fitting results of the HOPS-182 accretion streamer and some derived parameters across different stellar mass.** Rows 2 to 11 provide the right ascension (RA) offset, declination (Dec) offset, line-of-sight (LOS) offset, RA velocity component, Dec velocity component, LOS velocity component, kinetic energy, potential energy, specific angular momentum, and infall timescale, respectively. Offsets are relative to the source center. Rows 12 to 15 list the overall gravitational force density $f_{grav}$, the local gravitational force density at the position where the magnetic field curvature reaches its maximum $f_{grav,max}$, the angular velocity $\omega$, and the ratio between the angular velocity and the magnetic field strength $\omega/B$, respectively.